On the photoinduced phase transition in $(GeTe)_n(Sb_2Te_3)_m$

S. M. Yakubenya
NRC "Kurchatov Institute", 123182, Moscow, Russia

A. S. Mishchenko

RIKEN Center for Emergent Matter Science (CEMS), 2-1 Hirosawa, Wako, Saitama, 351-0198, Japan
NRC "Kurchatov Institute", 123182, Moscow, Russia

Abstract. We suggest a phenomenological description of the photo-conversion in Ge-Sb-Te phase-change memory alloys from amorphous to crystalline phase which explains why both photo-excitation and high temperatures $T > 160C$ are required for the transition from hexagonal to tetrahedral phase. The position of chemical potential at high temperatures allows light induced inverse population of the nucleons of the crystalline phase which are not stable otherwise. Then, inverse population accumulates holes on neighboring Te and Ge ions and locks the photo-conversion transition by pushing Ge ions into the interstitial position to minimize the Coulomb repulsion energy.

1. Introduction

In spite of the impressive number of studies devoted to Ge-Se-Te memory alloys there is no adopted agreement on the nature of photo-induced transitions in these materials. An idea to use the photo-induced transitions for data storage was put forward long time ago [1] and intense studies were followed by development of the technologies used in digital versatile discs (DVDs) and many other commercial devices [2]. Te-based chalcogenides is a class of pseudobinary compounds $(GeTe)_n(Sb_2Te_3)_m$ where the most promising for industrial applications, and, therefore, most intensively studied is $Ge_2Sb_2Te_5$ (GST). The latter is the most important for technologies because it stands over a million of cycles of fast switches between amorphous (hexagonal) and crystalline (tetrahedral) phases and demonstrates a very high stability of the photo-converted state.

Although GST is studied well enough to develop various industrial applications the knowledge of this material is still of purely empirical nature because enormous amount of accumulated experimental data [3, 4, 5] have not been resulted in appearance of any adopted opinion on the mechanism of the photo-induced phase transition. A lack of clear, at least phenomenological, picture giving clear clue to the nature of stabilization of the photo-converted phase hinders search for another classes of materials with similar or even more promising properties for electronics.

In this paper we suggest a phenomenological model which offers a mechanism of the stabilization of the photo-induced phase of GST-like materials. Suggested phenomenological picture is essentially based on the concept of double-defects [6, 7] which goes beyond the standard theory [8, 9] neglecting correlations between different types of defects. Double defect approach is obligatory generalization when the properties of a material are substantially dependent on the inter-defect interplay.

We introduce related to GST particular concepts of double-defect model in section 2, outline the equilibrium properties of the structures separated by the photo-induced transition in section 3, suggest a phenomenological mechanism of the stabilization of the photo-converted state in section 4, and present some final remarks in section 6.

## 2. Double defects in GST

The simplest description of defects in semiconductors considers every defect as isolated one [8, 9], e.g. interstitial $\{X_i\}$ (impurity or native ion in interstitial) or substitution $\{X_s\}$ (impurity or native ions in a lattice site) defect. In contrast, double-defect analysis [6, 7] considers the interstitial and substitution defects as two limiting cases of a double defect while the general object describing the system is a pair of interstitial $\{X_i\}$ and host vacancy $\{V_{host}\}$. Then, interstitial (substitution) defect is realized when the distance between counterparts is infinity (zero). Indeed, the host ion in such classification is a particular case of a substitution defect.

Particular type of double-defect related to GST is a pair defect where ion is moved from its lattice position into the location of the nearest interstitial. In particular, it is $\{Ge_i - V_{Ge}\}$ defect which consists of Ge ion in the interstitial position $Ge_i$ and Ge vacancy $V_{Ge}$. It has been shown that there are three kinds of electrons with different properties which can be considered as associated with this defect. These three kinds are (i) binding valence bond electrons; (ii) core state electrons; and (iii) dangling host hybrids [6, 7].

Binding valence bond electrons occupy binding valence bonds. There are two such electrons in the specific case of $\{Ge_i - V_{Ge}\}$ defect and these electrons saturate one valence bond. This type of electrons belong to the collective states of the valence band and the filling of these states weakly depends on the position of the chemical potential in the gap of semiconductor.

Another electrons whose filling which is insensitive to the position of the chemical potential are the core states. The anti-bonding component of these states is higher in energy than the vacuum level and, thus, binding component cannot exist in the energy scale which can be associated with states forming the electronic structure of the solids. It can be shown that any state lying below some critical energy $E_{cr}$, whose value is specific for particular compound and defect, belogs to core states [6, 7]. The value of critical energy for $\{Ge_i - V_{Ge}\}$ is larger than the second ionization potential ($\sim 16$ eV) of Ge [10] and smaller than the third ionization potential of manganese ($\sim 33$ eV) [10]. Such estimate for $E_{cr}$ is in agreement with data on the magnetization of $Ge_{1-x}Mn_xTe$ [11, 12] which show that the manganese ions are in the state $Mn^{+2}$ at $x \to 0$. Hence, the maximal charge state of such defect is $\{Ge_i - V_{Ge}\}^{2+}$ regardless of the position of the chemical potential $\mu$.

The third kind of electrons occupy orbitals which do not find another electron to create a valence bond but, at the same time, their energy is in the scale of typical energies of the band structure of semiconductor. In case of a pure interstitial defect such orbitals are called Dangling Bond Hybrids (DBH) [13]. Pair defect {$Ge_i$ - $V_{Ge}$} is the result of a jump of a host ion into neighboring interstitial position and we introduce the name Dangling Host Hybrids (DHH) for dangling electrons of pair defect. DHH are the only orbitals whose filling essentially depends on the position of chemical potential and, as we show in the next sections, are the key players in the physics of the photo-conversion.

## 3. Equilibrium properties of GST

There are several structural modifications $Ge_2Sb_2Te_5$ which are stable or metastable at different temperatures [14]. Typically, the film is deposited at room temperature by magnetron sputtering either on glass slides or on crystalline silicon covered with a native oxide [15]. The phase obtained at such conditions is amorphous (hexagonal) direct wide-gap semiconductor [16, 17] (see Fig. 1, right) with large $E_{og} \approx 0.7 \div 0.85$eV optical gap having large resistivity and small photoconductivity [15]. Amorphous phase has hole type of carriers arising from defects which are either interstitials or vacancies of cationic sublattice of Ge and Sb having concentration ranging from 2% to 20% [18]. A large concentration of defects and shallow localized in-gap states of DBH or DHH types, located within 80 meV from the valence band, manifests itself in Urbach tail observed in optical absorption spectra [15, 17]. According to Hall measurements the concentration of holes in an amorphous phase of similar compound [16] GeTe is around $3 * 10^{17} \div 5 * 10^{17}$cm$^{-3}$ which has to be roughly similar to concentration of holes in GST having very high value of resistivity which is close to that observed in GeTe [17].

Basing on the hole type of carriers one can estimate the location of chemical potential with respect to the valence band. Density of states in the valence band at temperature T is

$$N_v = 2\left(\frac{2\pi m_h k_B T}{(2\pi h)^2}\right)^{3/2} \qquad (1)$$

$k_B$ and h are Boltzmann and Planck constants, respectively, and $m_h \sim 1.15 m_0$ is effective mass of holes in GeTe [18] , $m_0$ is free electron mass. Resulting value at room temperature is $N_v \approx 4 \times 10^{19}$ cm$^{-3}$. Hence, the semiconductor in nondegenerate and holes concentration is given by the following expression

$$p = N_v \exp\left(\frac{E_v - \mu}{k_B T}\right) \qquad (2)$$

Then, position of the chemical potential is

$$\mu = E_v + k_B T \ln\left(\frac{N_v}{p}\right) \qquad (3)$$

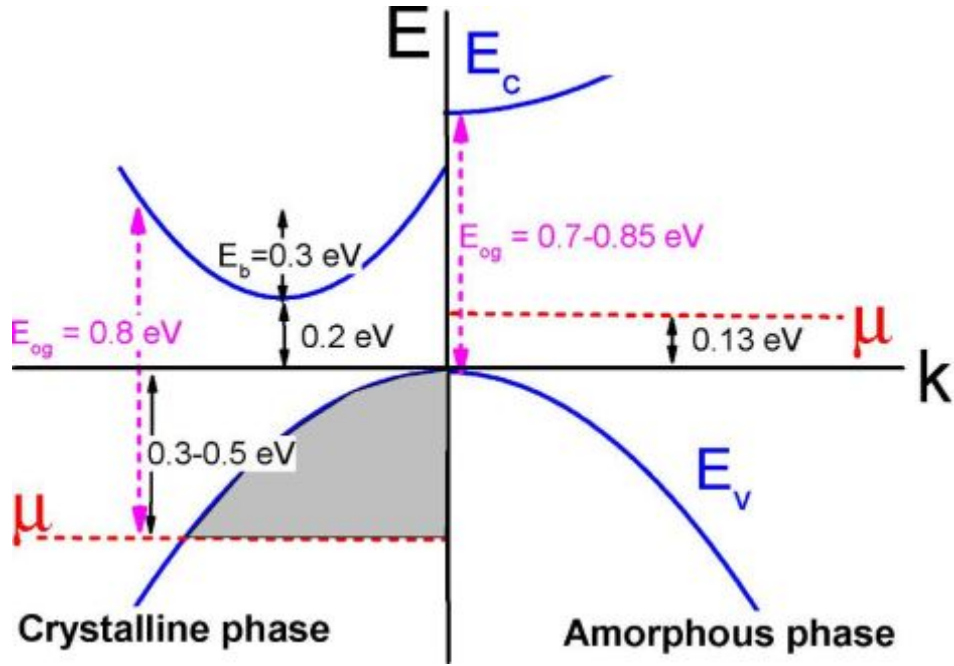

Figure 1. Schemsatic band structure of GST in amorphous (hexagonal) (right) and crystalline (tetrahedral) (left) phases. Solid blue lines show the dispersions of valence $E_v$ and conduction $E_c$ bands. Symbol $E_{og}$ stands for optical gap and $E_b$ is the Burshtein shift. Dashed red lines denote the positions of chemical potential $\mu$ at room temperature.

which is equal at room temperature to $\mu = E_v + 0.13$eV. Hence, free holes in the valence band originate from the thermal population of localized acceptor states of DBH and DHH orbitals near the top of the valence band. Chemical potential increases with temperature and equal to $\mu = E_v + 0.23$eV at T $\approx 130 \div 160$C which is the upper limit of the stability of amorphous phase in photo-conversion experiment.

The stable phase at the other side of the photo-conversion transition, i.e. at T $\geq 130 \div 160$C [16, 19] is the crystalline phase having distorted rock-salt structure [19]. Optical absorption experiments reveal direct gap in the range $0.73 \div 0.95$ eV depending on the holes concentration $1.8 \times 10^{20} \div 1.5 \times 10^{21}$ cm$^{-3}$, which seems to be very similar to that in amorphous phase. However, in spite of very close values of direct optical gaps in amorphous (hexagonal) and crystalline (tetrahedral) phases the band structures of these two phases are substantially different [15, 19]. The gap in amorphous phase is direct one with both minimum of conduction band and maximum of valence band located in the r-point of the Brillouin zone. To the contrary, the minimum of conduction band in the tetrahedral phase is shifted to the L-point and the indirect gap between the minimum of the conduction band in r point and the maximum of the valence band in L-point is 0.2 eV (see Fig. 1, left). It is shown that the chemical potential in crystalline phase is 0.3 ÷ 0.5 eV below the top of valence band [19]. Hence, the large optical gap of tetra-phase is a sum of (i) a energy separation of $\mu$ from the top of the valence band 0.3 ÷ 0.5 eV; (ii) indirect gap 0.2 eV; and (iii) Burshtein shift shift of chemical potential [20].

It is important to emphasize that both counterparts of the phase transition, amorphous and crystalline phases, have the same unique energy reference point set by the top of the valence band which is dominated mainly by p-orbitals of Te [21]. Since two phases differ mostly in the distribution of the ions of cationic sublattice, whereas the Te-sublattice does not change [22, 23], the energies of the top of the valence bands $E_v$ with respect to vacuum level are the same in amorphous and crystalline phases. Experimental data indicate that the amorphous phase is thermally excited into the tetrahedral phase even without photo-excitation. Such phenomenon occurs due to thermal excitation of carriers through the semiconducting gap $E_0 \approx$ 0.8eV. It is clear from band structure and known from experiment that the crystalline phase has very low and temperature independent resistivity $\rho_{cr}$ whereas that of the amorphous phase is very large. On the other hand, measurements of the temperature dependent $\rho_{am}(T)$ resistivity of amorphous thin films of GeTe shows that the ratio $\rho_{am}(T)/\rho_{cr}$ obeys the following equation

$$\frac{\rho_{am}(T)}{\rho_{cr}} = \exp\left(\frac{E_0/2}{k_B T}\right) \qquad (4)$$

in the very wide range of temperatures 77-410K [16]. Similar result was obtained in GST225 and GST124 with corresponding activation energies $E_0 = 0.42$eV and $E_0 = 0.45$eV, respectively. Such behavior can be explained if one assumes that highly conductive thermally activated nucleons of crystalline phase with high conductivity are populated by thermally activated transitions from the amorphous phase with low conductivity. These conductive nucleons represent the leading channel of conductivity which is the leading contribution to the conductivity on top of low-conductive amorphous phase.

## 4. Photoinduced transition in GST

We note that the thermal excitation of crystalline high conductivity nucleons does not convert the amorphous phase into the crystalline one because the nucleons are unstable in the situation when the concentration of the holes is small. To the contrary, in case of generation of large amount of carriers by photoexcitation through the semiconducting gap or by high density current the crystalline phase stabilizes [16, 19]. Note, both excessive nonequilibrium carriers and high enough temperature T > 160C are required for photo-conversion.

Our phenomenological description of the photo-conversion is the following. The characteristic time of of the inter-band recombination processes is much longer than that required for intra-band relaxation. Electronic intra-band relaxation time in GST is around $10^{-15}$s which is only two orders of magnitude faster than the typical lattice relaxation time which is set by typical phonon oscillation period $10^{-13}$s. On the other hand, two characteristic inter-band recombination times, which can be attributed to holes and electrons, are many orders of magnitude slower, $1.35 \times 10^{-3}$s and $4.25 \times 10^{-3}$s [16] respectively. Hence, there is a plenty of time to create inverse population and to pump significant concentration of electron-hole pairs which, in turn, have enough time to be equilibrated by intra-band and lattice relaxation before recombination processes with long characteristic times can return the system to equilibrium.

The elevated temperature T > 160C is required for photo-conversion because the chemical potential has to be put into the position which allows to start initial inverse population of the bands in the nucleons of the crystalline phase. As we mentioned before, the top of the valence band $E_v$ sets up the absolute energy reference point for both crystalline and amorphous phases. Hence, one can use the top of the valence band to compare positions of chemical potential $\mu$ in the amorphous phase and in the nucleon of the crystalline phase. We note that the temperature T > 160C is high enough to push chemical potential of the amorphous phase $\mu = E_v + 0.23$eV above the bottom of conduction band $E_c^{bottom} = E_v + 0.20$eV of the nucleon of crystalline phase. It is natural to assume that such position of the chemical potential with respect to photo-converted phase conduction minimum allows to stabilize the initial inverse population by formation of nucleons of crystalline phase.

As the result of inverse population there is significant number of holes and electrons in the valence and conduction bands, respectively. However, the properties of excess holes and electrons are different. Excess electrons are located on the anti-bonding states of the conduction band and, hence, are well delocalized in space. On the other hand, the top of the valence band and localized states above it are dominated by the bonding states of Ge and Te and, hence, large concentration of long-living holes leads to excess positive charges located on the neighboring sites occupied by Ge and Te ions. The second our assumption is the following. Strong repulsion of neighboring Ge and Te ions of leads to creation of pair defects where positively charged Ge ion is transferred into interstitial position to minimize the energy of Coulomb repulsion from the positively charged Te ion. We note that our second assumption is in close agreement with location of the chemical potential in the photo-converted phase. According to the double defect model the transfer of the Ge ion from lattice position with coordination number equal to 6 into the interstitial position with that equal to 4 leads to decrease of valence electrons by two because two of s-electrons does not longer participate in valence bonds. Therefore, transition of the Ge ion into interstitial adds two holes and the state of the pair defect can be described as

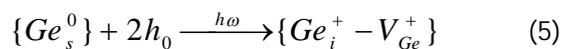

$$\{Ge_s^0\} + 2h_0 \xrightarrow{\ h\omega\ } \{Ge_i^+ - V_{Ge}^+\} \qquad (5)$$

where $h_0$ is a hole in the valence band. Note, that measured concentration of holes in crystalline phase p = $3 * 10^{20}$cm$^{-3}$ at T=160C is in good agreement with position of the chemical potential in this phase. Using Eqs. (1-3) on gets $\mu \approx E_v - 0.5$eV which is close to values measured in optical experiments [19]. Finally, stabilization of chemical potential deep in the valence band prevents pair defect from annihilation because both Ge and Te ions are positively charged. We note that the transition (5) is more effective in case when the single-hole occupation of of DHH orbitals is unstable due to formation of Anderson negative-U system [24].

## 5. Conclusions

We suggest a phenomenological model of the amorphous to crystalline phase photo- conversion of GST. The model utilizes unique interrelation between the band structures of the counterparts of the phase transition and specific difference of lattice site position of Ge in amorphous material and it's interstitial position in the crystalline phase. We show that both inter-gap excitation of carriers and elevated temperature T > 160C are required for photo-conversion because the inverse population can grow only when the chemical potential is located in the conduction band of the nucleons of the photo- converted crystalline phase. We assumed that Coulomb repulsion in the material with inverse electron-hole population can create pair defects which intensively create extra holes in the valence band and showed that such assumption is consistent with position of the chemical potential potential of crystalline phase. Massive creation of extra holes locks the photo-conversion transition pushing the chemical potential deep inside the valence band.

## 6. Acknowledgments

ASM was funded by the ImPACT Program of the Council for Science, Technology and Innovation (Cabinet Office, Government of Japan).

## 7. Literature

[1] Ovshinsky S R  *Phys. Rev. Lett.* 21 ,1450,1968
[2] Ohta T  *J. Optoelectron. Adv. Matter.* 3, 609,2001
[3] Kolobov A V, Fons P, Frenkel A I, Ankudinov A L,Tominaga J, and Uruga T , *Nature Mater. 3 ,*703,2004
[4] Atta-Fynn R, Biswas P, Drabold D A  *Phys. Rev.B* 69, 245204,2004
[5] Ohta T and Ovshinsky S R  *Photo-induced Metastability in Amorphous Semiconductors,* (Wiley-VCH, Weinheim. )  2003,310 p.
[6] Yakubenya S M , *Solid State Physic* 33, 824,1991
[7] Yakubenya S M, , *Solid State Physic* 33, 829,1991
[8] Bonch-Bruevich V L , Kalashnikov S G 1990 *Semiconductors physics* , Nauka (in Russian) , 688p.
[9] Fistul' V I, 1969 *Heawilly doped semiconductors,* Plenum Press (New York), 397 p.
[10] Radtsig A A , Smirnov B M 1986 2nd ed.*Handbook Parameters of atoms and atomic ions* , Energoatomizdat, (in Russian), 2nd ed.(1986) 344p.
[11] Ciucivara A, Sahu B R, abnd Klenman L, *Phys. Rev. B* 75, 241201(R), 2007
[12] Cochrane R W, Plischke, and Strom-Olsen J O, *Phys. Rev. B* 9, 4013, 1974
[13] Zanger A ,Lindefelt U  *Phys.Rev. B 27 ,*1991, 1983
[14] Yamada N, Ohno E, Nishiuchi K, Akahira N, and Takao M , *J. Appl. Phys.* 69, 2849,1991
[15] Lee B-S , Abelson J R, Bishop S G, Kang D-H, Cheong B-K, and Kim K-B , *J. Appl. Phys.* 97, 093509, 2005
[16] Bahl S K and Chopra K L , *J. Appl. Phys.*41, 2196,1970
[17] Xu L, Tong L, Geng L, Yang F, Xu J, Su W, Liu D, Ma Z and K. Chen K, J. Appl. *Phys.* 110 , 013703,2011

[18] Kikoin A K 1976 *Handbook Physical values tables,* Atomizdat (in Russian), 1008p.

[19] Bahl S K and Chopra K L 1969 *J. Appl. Phys.* 40 , 4940

[20] Moss T.S., *Optical properties of semiconductors* (Butterworths Scientific Publication, London , 1959)

[21] Singh D J 2013 , *J. Appl. Phys.* 113, 203101

[22] Fons P, Osawa H, Kolobov A V, Fukaya T, Suzuki M, Uruga T, Kawamura N, Tanida H, and J. Tominaga J 2010, *Phys. Rev. B* 82, 041203(R)

[23] Jung M-C, Shin H J, Kim K, Noh J S and Chun J 2006, *Appl. Phys. Lett.* 89, 043503

[24] Baraff G. A.,Kane E. O.,and Schliiter M. 1980,Phys. *Rev. B* 21, 5662